\RequirePackage{snapshot}

\documentclass[9pt,twocolumn,twoside]{osajnl}

\journal{josab} 

\setboolean{shortarticle}{false} 
\usepackage[caption=false]{subfig}
\usepackage[export]{adjustbox}

\title{Carrier-envelope phase effects in graphene}

\author[1,2,*]{C. Lefebvre}
\author[1,2]{D. Gagnon}
\author[1,2]{F. Fillion-Gourdeau}
\author[1,2]{S. MacLean}

\affil[1]{INRS-EMT, 1650 Lionel-Boulet, Varennes, J3X 1S2 Canada}
\affil[2]{Institute for Quantum Computing, University of Waterloo, Waterloo, Ontario, N2L 3G1 Canada}

\affil[*]{Corresponding author: catherine.lefebvre@emt.inrs.ca}

\ociscodes{(260.3160) Physical optics, Interference; (320.5540) Ultrafast optics, Pulse shaping; (260.3230) Physical optics, Ionization; (020.5580) Atomic and molecular physics, Quantum electrodynamics; (160.4330) Materials, Nonlinear optical materials}

\begin{abstract}
We numerically study   the interaction of a terahertz  pulse with monolayer graphene. We observe that the electron momentum density is  affected by the carrier-envelope phase  (CEP) of the single to few-cycle terahertz laser pulse that induces the electron dynamics. In particular, we see strong asymmetric electron momentum distributions for non-zero values of the CEP. We explain the origin of the asymmetry within the adiabatic-impulse model by finding conditions to reach minimal adiabatic gap between the valence band and the conduction band. We discuss how these conditions and the interference pattern, emanating from successive non-adiabatic transitions at this minimal gap, affect the electron momentum density and how they are modified by the CEP. 
This opens the door to control  fundamental time-dependent electron dynamics in the tunneling regime in Dirac materials. Also, this control suggests a way to measure the CEP  of a terahertz laser  pulse  when it interacts with condensed matter systems.
\end{abstract}

\setboolean{displaycopyright}{true}

\begin{document}

\maketitle

\section{Introduction}

When strong laser pulses of ultrashort duration interact with matter, the pulse shape affects the nonlinear response of the physical system. The laser electric field 
 is characterized by the pulse envelope, 
 the field amplitude, 
 the carrier wave frequency 
 and the carrier-envelope phase (CEP). The latter is defined as the phase between the carrier wave and the position of the pulse envelope maximum amplitude.
 In particular, the  CEP plays a major role in the dynamics of physical systems when the pulse duration contains only  a few optical cycles of the carrier frequency.  
 
The effects of the CEP has been demonstrated in the nonlinear ionization of atoms \cite{PhysRevA.65.061802,PhysRevLett.91.253004}  and molecules \cite{PhysRevLett.94.203003,Kling246} in gas phase. 
For instance, during the dissociative ionization of deuterium molecules induced by a few-cycle femtosecond laser pulse, the proper selection of the CEP results in the localization of the electron on one of the two D$^+$ ions \cite{Kling246}. This CEP-controlled asymmetric fragmentation reveals the importance of the ultrashort laser pulse waveform on the electron dynamics. This study  has been extended to more complex molecular systems, such as CO \cite{PhysRevLett.103.103002}, as well as to semiconductors \cite{Schiffrin_Nature493_70}, metals \cite{Kling_APL2_036104} and nanostructured materials \cite{StockmanApplPhysA89_247}.  

In condensed matter, 
it is the collective electron motion of the material rather than the electron itself that is controlled by the CEP of the few-cycle laser pulse \cite{Zherebstov_NatPhys7_656}. 
A particular example is graphene, which is a unique two-dimensional (2D) material that exhibits perfect delocalization of the electrons in a single layer of carbon atoms that have arranged hexagonal periodicity \cite{Geim_NatureMat6_183,RevModPhys.81.109}.  Graphene has a relativistic-like dispersion relation, no mass gap and is characterized by a Fermi velocity that obeys $v_F \approx c/300$, where $c$ is the speed of light.
 The charge transport in graphene can be described, in the low-energy limit of the tight-binding model, by an effective theory based on a massless 2D Dirac equation. These properties make graphene, and other Dirac materials \cite{doi:10.1080/00018732.2014.927109}, interesting simulators to explore  relativistic-like electron dynamics. 
 In fact, in analogy to the nonperturbative quantum electrodynamics (QED) process of electron-positron pair production  \cite{PhysRev.82.664}, electron-hole pairs are generated in the tunneling regime in Dirac materials \cite{PhysRevD.78.096009}. This relativistic-like regime in condensed matter is reached with accessible laser sources at much lower laser intensities than required in QED  \cite{RevModPhys.84.1177}. Owing to their massless nature,   Dirac materials do not suffer from an exponential suppression of the pair rate as it is the case in QED \cite{PhysRevB.94.125423}.
Theoretical approaches described the carrier dynamics in graphene and its nonlinear electromagnetic response induced by an optical \cite{PhysRevB.84.205406,Jafari_JPCondMatt24,PhysRevB.95.035405} or THz \cite{Mikhailov_EPL79_27002,PhysRevB.90.245423,PhysRevB.93.085403} radiation. 

The ultrafast dynamics of electrons in graphene has been studied recently in terms of electron-electron interactions (e.g. Auger processes) via laser pump-probe spectroscopy which involves ultrashort pulses of few femtoseconds duration \cite{Brida:2013kn}. It has also been shown that electronic properties of graphene evolve on an ultrashort time scale. For instance,  the time-evolution  of the electron and hole quantum distribution is in the range of 500 fs \cite{Gilberston_JPhysChemLett3_64}. In the absence of scattering, significant electron transfer from the valence band to  the conduction band can persist after the end of the laser pulse, generating a current that is much higher than that in dielectrics or metals \cite{PhysRevB.91.045439}. Not only can this current originating from graphene's unique electronic properties be used for technological development in optoelectronics, the properties of ultrashort laser pulses initiating the electron dynamics may be optimized. In particular, the CEP of ultrashort laser pulses can play a role in the non-adiabatic interband dynamics.
The influence of the CEP-induced temporal variations of the laser field on the interband transitions and generated current has been reported theoretically with  terahertz radiation pulses of one to few cycles duration \cite{1367-2630-15-5-055021}.  Experiments combined with a theoretical interpretation demonstrated the control of the residual conducting current by light-field waveform shaped by the CEP in the near-infrared spectral range  \cite{Higuchi:2017fk}.

In this paper, we are interested in the CEP effects on the electron-hole pair production in graphene subjected to a single to a few-cycle THz laser pulse. Specifically, we focus on the electron momentum distribution (EMD) after the laser excitation and we observe an asymmetry in the distribution while varying the CEP. Using the adiabatic-impulse model for interpretation, we isolate the origin of the asymmetry controlled by the CEP.
The ultrafast electron dynamics in graphene subjected to THz radiation is described by an adiabatic evolution in the valence and conduction bands, followed by successive non-adiabatic transitions. These successive non-adiabatic transitions give rise to quantum interferences.
Also, interband transitions occur only when the mass gap is minimal. In turn, the amplitude of the transition probability depends on the instantaneous field amplitude, controlled by the waveform and hence, the CEP. 
%
This is a very selective mechanism that can be used for  the enhancement of optical phenomena and temporal control of carrier dynamics in graphene with CEP-tunable single to few-cycle laser pulses. 
Ultimately, the observed strong sensitivity of the momentum pair distribution could be used to determine the CEP of THz single to few-cycle laser pulses in interaction with 2D materials. 

We choose to study the interaction of graphene with THz radiation because, in contrast to the near-infrared regime, THz radiation can drive electron tunneling in materials without inducing thermal effects \cite{Yoshioka:2016jk}. This is well-suited for the study of electron-hole pair production in graphene. As the THz pulse duration can easily be reduced down to few optical cycles (sub-picoseconds duration), the CEP of the pulse needs to be characterized and tuned  \cite{Blanchard:07,Yoshioka:2016jk}. 
Studies of CEP measurement have been performed in atomic, molecular and optical physics  mostly using few-cycle near-infrared laser pulses (at $\lambda=800$~nm). These methods are based for instance on $f$-to-$2f$ interferometry \cite{Jones_Science288_635}, which averages over a large number of CEP-stabilized laser shots, or on photoelectrons generated by above-threshold ionization  \cite{Wittmann:2009fk}, which performs single-shot CEP measurements of either CEP-stabilized or non-stabilized pulses (i.e. CEP tagging).
In principle, the non-linear generation of THz radiation implies CEP locked pulses, but it does not allow for a direct change of the CEP itself. The tunability of the CEP THz radiation has only recently been suggested via Gouy phase shift manipulation \cite{Yoshioka:2016jk}. 
The control and the tunability of the CEP of the THz radiation opens the door to time-controlled dynamics and deeper understanding of time-dependent processes involving  relativistic-like electrons in Dirac materials. Our contribution emphasizes the critical role of the CEP of a THz radiation in the ultrafast electron dynamics in monolayer graphene.

This paper is organized as follows. In Section \ref{section_simulation}, the method used for the numerical simulations is provided. Then the numerical results are presented and discussed in Section \ref{section_ResultsDiscussion}, which demonstrate the effects of the CEP of a 10 THz laser pulse on the electron momentum density of graphene. Finally, concluding remarks are given in Section \ref{section_conclusion}.

\section{Simulations}\label{section_simulation}

We consider the interaction of a homogeneous THz radiation with monolayer graphene. 
We also consider that the momentum of the quasi-particles is close to a Dirac point, ensuring a linear dispersion relation and a physical description in terms of the Dirac equation. 
Under these assumptions, the Fermion dynamics is described by the 2D time-dependent massless Dirac equation in momentum space (unless specified, Lorentz-Heavyside natural units are used where $\hbar=c=1$ and the electron charge $e=\sqrt{4\pi\alpha}$, with $\alpha$ the fine-structure constant): 
\begin{equation}\label{eq_Dirac}
i\partial_t \psi_{s,\textbf{K}_{\pm}}(t,\textbf{p}) = H_{\textbf{K}_{\pm}}  (t,\textbf{p}) \psi_{s,\textbf{K}_{\pm}}(t,\textbf{p}),
\end{equation}
with the time-dependent two-spinor wavefunction  $\psi_{s,\textbf{K}_{\pm}}(t,\textbf{p})$ and the physical spin of the electron $s=\pm$.
  Within the first Brillouin zone of graphene, there are two non-equivalent Dirac points labeled $\textbf{K}_{+}$ and $\textbf{K}_{-}$.  These are points in the dispersion relation where the conduction and valence bands meet. They are positioned at the absolute momentum $|\textbf{K}_{\pm}|=\frac{4\pi}{3\sqrt{3a}}\approx 3361$ eV, where $a\approx 1.42\times10^{-10}$ m is the distance between carbon atoms. Then, 
 $\textbf{p} = (p_{x},p_{y})$ is the relative momentum around non-equivalent Dirac point. 
The Hamiltonian is
\begin{equation}\label{eq_Hamiltonian}
H_{\textbf{K}_{\pm}} (t,\textbf{p}) = \pm v_{F} \boldsymbol{\alpha} \cdot (\textbf{p}+e\textbf{A}(t)).
\end{equation}
$\textbf{A}(t)$ is the time-dependent vector potential related to the electric field, i.e.  $\textbf{E}(t)=-\partial_t \textbf{A}(t)$ and $\boldsymbol{\alpha} = (\alpha_{x},\alpha_{y})$ are the Pauli matrices.

In this article, the main observable considered is the electron momentum density (EMD) $f_{s,\mathbf{K}_{\pm}}(\mathbf{p})$, giving the (dimensionless) density of positive charge carriers in phase space.
In principle, probing the EMD observable can be achieved using time- and angle-resolved photoemission spectroscopy (tr-ARPES) \cite{PhysRevB.93.155434, PhysRevB.96.020301}. 
Alternatively, if one is interested in measuring the integrated EMD, for example to probe its anisotropy, measuring the photo-current is a possibility \cite{1367-2630-15-5-055021, Higuchi:2017fk}.
In all calculations presented in this article, we assume that the Fermi energy is equal to zero, and that the temperature is sufficiently small to prevent thermal transitions from the valence band to the conduction band.
In other words, the valence band is assumed to be completely filled up to the Fermi level before the application of an external field.
This means that only photo-excitation contributes to the EMD observable as the THz pulse interacts with graphene. 
With all these considerations, the EMD can be calculated from the number operator in the ``in/out'' formalism. After some simplifications, the result of this procedure is \cite{PhysRevB.92.035401,FillionGourdeauRussianJ2017,PhysRevB.94.125423}
\begin{equation}\label{eq_pairDensity}
f_{s,\mathbf{K}_{\pm}}(\mathbf{p}) = \frac{1}{2E_{\textbf{p}}2E_{\textbf{p}}}\arrowvert u_{s,\textbf{K}_{\pm}}^{\dagger}(\textbf{p}) \psi_{s,\textbf{K}_{\pm}}(t_f,\textbf{p})\arrowvert ^2,
\end{equation}
along with the following initial condition on the wave function
\begin{equation}
\psi_{s,\textbf{K}_{\pm}}(t_i,\textbf{p}) = v_{s,\textbf{K}_{\pm}}(-\mathbf{p})
\end{equation}
and where $E_{\textbf{p}} = v_{F}|\mathbf{p}|$ is the energy. In other words, the EMD is related to the probability of transition from the negative energy states (conduction band) to the positive energy states (valence band) under the effect of the electric field. This is evaluated by preparing a negative energy state $v_{s,\textbf{K}_{\pm}}(-\mathbf{p})$ at the initial time $t_i$ and by propagating numerically up to the final time $t_f$ using \eqref{eq_Dirac}. The propagation is performed using a second order split-operator decomposition scheme \cite{PhysRevB.92.035401,FillionGourdeauRussianJ2017,PhysRevB.94.125423}.
The final state is then projected onto a free positive energy state $u_{s,\textbf{K}_{\pm}}^{\dag}$. The explicit expression of the positive and negative energy states are given in Appendix \ref{app:free_states}.

We note here that in the calculation of the EMD, it is assumed that the asymptotic value of the vector potential is zero, that is $\mathbf{A}(t)=0$ for $t \in (-\infty,t_{i}] \cup [t_f,\infty)$. Physically, this implies that there is no DC component in the electric field. If this condition is not fulfilled, the expression of free states and energies has to be modified to preserve gauge invariance. We refer the reader to Ref. \cite{PhysRevB.92.035401} for more details on this issue.

For the purposes of this article, the electric field is assumed to be homogeneous and linearly polarized along the $x$ axis:
\begin{equation}\label{eq_field}
\textbf{E}(t)=-\partial_t \textbf{A}(t)= -\hat{x} \partial_t A_x (t) .
\end{equation}
The vector potential is defined as
\begin{equation}
A_x(t) = - \frac{E_0}{\omega} \epsilon_0(t) \sin(\omega t + \phi)
\end{equation}
where $E_0$ is the maximum amplitude of the electric field, $\omega$ and $\phi$ are the carrier-wave frequency and the CEP, respectively, and $\epsilon_0 (t) \in [0,1]$ is an  envelope function.
In this article, we use the following envelope \cite{PhysRevA.65.061802}:
\begin{equation*}\label{eq_env}
\epsilon_0 (t) = 
\begin{cases}
  0, & \mathrm{for}\ t\notin \left[0,NT \right]\\
  \sin^2(\frac{\pi t}{NT}), & \mathrm{for}\ t\in \left[0,NT \right],
\end{cases}
\end{equation*}
with $T=2\pi/\omega$ the period of the carrier-wave and $N$ the total number of optical cycles under the pulse envelope.
This choice of vector potential ensures there is no DC component \cite{PhysRevA.65.061802}.

Examples of the electric field, $E(t):= E_{x}(t)$ (black line) and vector potential, $A(t):=A_{x}(t)$ (green line) are depicted in the right panels of Fig. \ref{fig_momentum_1cycle}   for three different values of CEP, $\phi=0, \pi/4, \pi/2$ and following  field parameters: frequency $\nu=10$ THz ($T=100$ fs),   field strength $E_0=10^7$ V/m (equivalent to 6.516  natural unit of electric field and a laser intensity of $I \propto  \frac{1}{2}E_0^2 = 2.6\times 10^{11}$ W/m$^2$) and  pulse duration of one period ($N=1$). The CEP modifies the waveform of the electric field under the envelope by shifting the maximum of $E(t)$ with respect to the center of the envelope. 
We note that for any pulse duration with integer number of $N \in \mathbb{N}$, the energy of the pulse under the envelope is the same for all values of $\phi$, such that  the total field energy is the same for any CEP value. Consequently, any enhancement or suppression of an observable with varying CEP is physical and due to the change of the waveform controlled by the CEP. 

Let us conclude this section by discussing experimental conditions under which the (idealized) Dirac equation model is reasonably applicable for THz pump pulses.
There are essentially three conditions to be met if a photo-induced momentum-space distribution is to be measurable: (i) graphene flakes must be large enough to prevent scattering at edges, (ii) phonon scattering should be small and (iii) the electron-electron coupling constant should be small.
Condition (i) may be satisfied with current monolayer graphene synthesis methods \cite{RevModPhys.86.959}, while condition (ii) may be satisfied using a low temperature apparatus \cite{PhysRevLett.99.086804}.
Condition (iii) on electron-electron scattering, however, is likely more difficult to achieve in the laboratory because of the associated thermalization time.
This is because, as detailed in Refs. \cite{ANDP:ANDP201700038, Oladyshkin2017} the largest reported thermalization times in graphene are a few hundreds of femtoseconds, which is shorter than the duration of a few-cycle THz pulse.
Consequently, the observability of anisotropic momentum-space distributions with THz pump beams implies experimental precautions to increase the thermalization time in graphene to the picosecond range \cite{PhysRevB.94.125423, arXiv:1710.09889}.
This includes performing experiments at temperatures below 10 K \cite{PhysRevB.90.245423} and using substrates which screen electron-electron interactions owing to their very high dielectric constant \cite{PhysRevLett.107.225501}.
It should be noted, however, that anisotropic momentum-space distributions have already been measured in pump-probe experiments in the optical regime in Ref. \cite{PhysRevB.96.020301}.
Similar measurements in the THz regime may thus very well happen in the near future.

The effect of electron-electron scattering, defect scattering and electron-phonon scattering on the time evolution of charge carriers can be included theoretically via the Graphene Bloch equations \cite{PhysRevB.84.205406,PhysRevB.90.245423}. In this article, as mentioned above, these effects are neglected.  

\section{Results and Discussion}\label{section_ResultsDiscussion}

\begin{figure*}[t!]
\centering
\includegraphics[height=12cm]{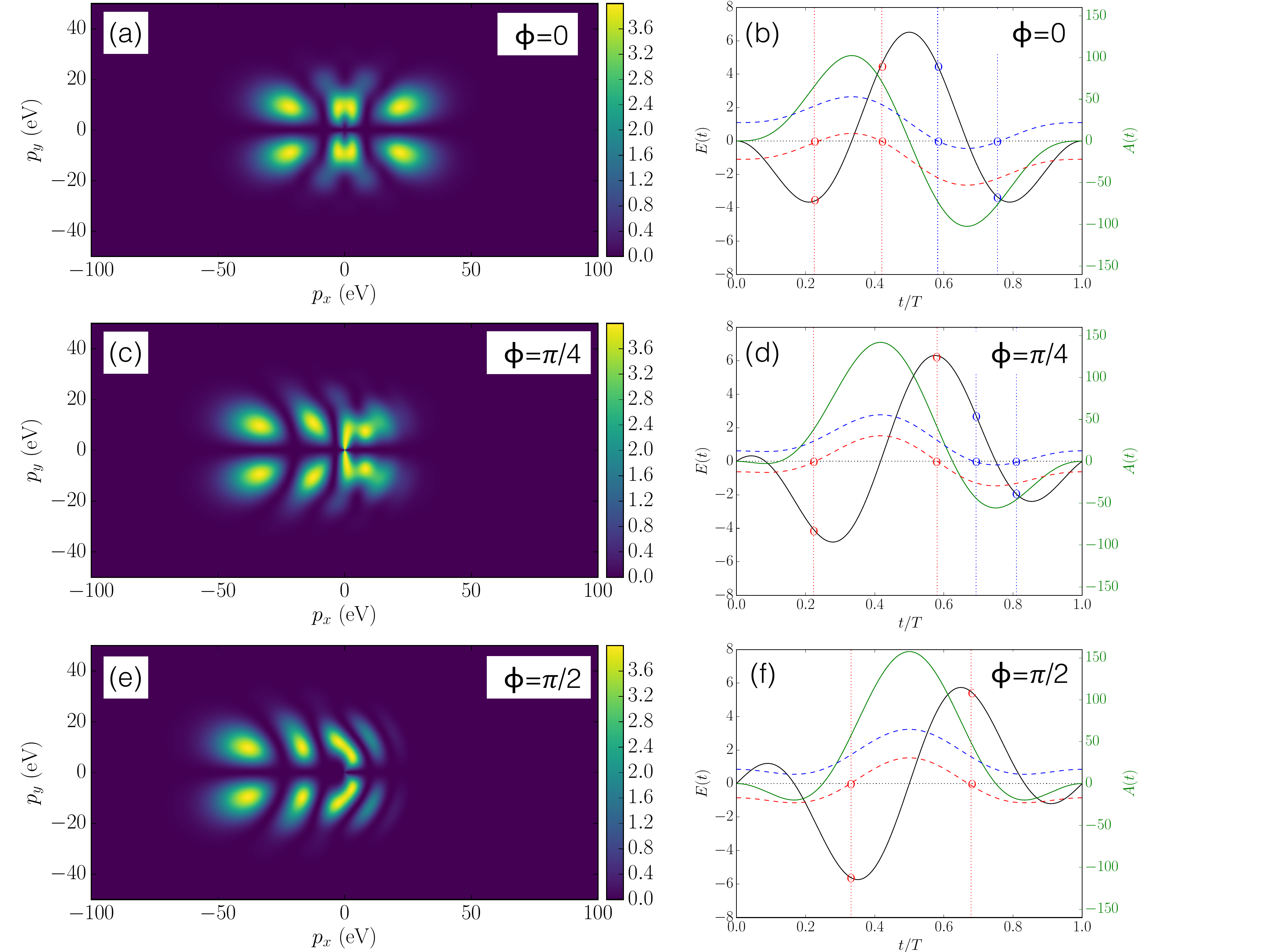}
        \caption{(a,c,e) Electron momentum density maps in graphene for three different values of CEP $\phi=0, \pi/4, \pi/2$.  (b,d,f) Associated electric field, $E(t)$ (black line) and vector potential, $A(t)$ (green line) with parameters: one optical cycle ($N=1$); 10 THz;  $10^7$ V/m. Also displayed is the time-evolution of $p_x + e A(t)$ with $p_x<0$ (red dashed line) and $p_x>0$ (blue dashed line) for representative values of (b) $|p_x|=22$ eV, (d) $|p_x|=12.5$ eV, (f) $|p_x|=17$ eV. The red/blue circles indicate the solutions of $p_x + e A(t) = 0$, i.e. times where Eq. (\ref{condition_zero}) is fulfilled and the non-adadiabatic transition probability $P_j$ is large.}\label{fig_momentum_1cycle}
\end{figure*}

The numerical results of the electron momentum density (EMD) induced by the interaction of a monolayer graphene sample with a one-cycle laser pulse of strength $E_0=10^7$ V/m, frequency $\nu=10$ THz are shown in Fig. \ref{fig_momentum_1cycle} (left panels). The electric field is linearly $x$-polarized and consequently, the momentum maps are symmetric along the transverse momentum $p_y$.  A change in the CEP  imparts an important modification on the waveform of the electric field $E(t)$ (right panels, black line) and its vector potential $A(t)$ (green lines). 
For $\phi=0$, the temporal shapes of the electric field $E(t)$ is symmetric with respect to $t=T/2$, the time when the envelope function is maximal. One observes that the symmetry carries over to the EMD map in the reciprocal space, which is symmetric  with respect to $p_x=0$.
However, for  $\phi \ne a\pi$, $a=0,1,2,3,\cdots$, e.g. $\phi=\pi/4$ and $\pi/2$ (panels (c)-(d) and (e)-(f)), the time-dependent electric field is shifted by the phase $\phi$ with respect to the maximum of the field envelope. This imparts  an asymmetry to the total field seen by the electrons.
 As a result, the EMD are  modified by the change of the temporal pulse shape controlled by the CEP and we observe an asymmetry in the longitudinal momentum $p_x$ (panels (c), (e)). 




\subsection{Adiabatic-impulse model}

Referring to the adiabatic-impulse model \cite{PhysRevB.94.125423,Shevchenko20101}, we will now see how the time shift imposed by the CEP influences the electronic wavefunction in the reciprocal space. The adiabatic-impulse model is defined in the velocity gauge and is used  to support the interpretation of the physical features observed in the full numerical calculations of the EMD. 
 In graphene subjected to an homogeneous field, the electron Hamiltonian is analogous to a two-level system. When the system is driven adiabatically, the quantum dynamics of the electron is characterized by successive adiabatic evolutions followed by non-adiabatic transitions. The highest probability of transition occurs when interband coupling is maximal, i.e. when the field amplitude has the largest value. However, the transition does not necessarily occur when the field is maximal. Rather, this takes place when the adiabatic mass gap is minimal, corresponding to the condition
 \begin{equation}\label{condition_zero}
p_x + e A(t) = 0.
\end{equation}
By solving this equation, one  finds the set of transition times $(t_{j})_{j=1,\cdots,2N}$, where $2N$ is the number of half-cycles. For instance, under a pulse envelope of duration of one cycle, this happens twice, at each half cycle (at $t_1$ and $t_{2}$).
In the adiabatic basis,  the population is initially in the adiabatic lower level (the valence band)  of the system as described by
\begin{equation}\label{Prob_init}
	B^{\mathrm{adiab}}(t_0)=
	\begin{pmatrix}
		0 \\ 1
	\end{pmatrix}.
\end{equation}
The population evolves adiabatically in each respective band according to the adiabatic time evolution operator
\begin{equation}\label{matrix_adiab}
	U^{\mathrm{adiab}}_{t_{j+1},t_j}=
	\begin{pmatrix}
		e^{-i\xi_{j+1,j}} & 0 \\ 0 & e^{i\xi_{j+1,j}}
	\end{pmatrix},
\end{equation}
where $\xi_{j+1,j}$ is the phase accumulated during the adiabatic evolution between $t_{j+1}$ and $t_j$ when  the system encounters an avoided crossing of the adiabatic energies near a Dirac point. At the avoided crossing, there is a probability 
\begin{equation}
P_{j}=e^{-\pi \frac{v_F p_y^2}{e|E_x(t_j)|}}, 
\end{equation}
that the  population is transferred non-adiabatically from the lower to the upper state (from the valence band  to the conduction band). We note that this probability is exponentially suppressed at higher values of the transverse momentum $p_y$.
The non-adiabatic transition matrix is defined as \cite{Shevchenko20101,PhysRevA.75.063414}
\begin{equation}\label{matrix_nonAdiab}
	N_j=
	\begin{pmatrix}
		\sqrt{1-P_j} e^{-i\varphi_j}& -\sqrt{P_j} \\ \sqrt{P_j} & \sqrt{1-P_j}e^{i\varphi_j}
	\end{pmatrix}
\end{equation}	
where $\varphi_j$ is the phase accumulated during each non-adiabatic transition, (the so-called Stokes phase).
Then the evolution in each band is adiabatic, following Eq. (\ref{matrix_adiab}) until the next avoided crossing is encountered at $t_{j+1}$  where another non-adiabatic interband transition occurs, governed by Eq. (\ref{matrix_nonAdiab}).

In our case, an envelope $\epsilon_0 (t)$ modulates the laser pulse, such that the field amplitude of the electric field at each half optical cycle is not equivalent, and as a consequence $P_{1}\ne P_{2}$. 
The wavefunction in the adiabatic basis at the end of the period of the laser pulse is expressed as 
\begin{eqnarray}\label{Prob_final_factor}
	B^{\mathrm{adiab}}(t_f)&=&U^{\mathrm{adiab}}_{t_{f},t_{2}}N_{2}U^{\mathrm{adiab}}_{t_{2},t_{1}}N_{1}U^{\mathrm{adiab}}_{t_{1},t_0} B^{\mathrm{adiab}}(t_0),\\ \nonumber
	&=&
	\begin{pmatrix}
	B^{\mathrm{adiab}}_{1}(t_f) \\
	B^{\mathrm{adiab}}_{2}(t_f)
	\end{pmatrix},
\end{eqnarray}
where
\begin{align}
B^{\mathrm{adiab}}_{1}(t_f)&=
		-\sqrt{P_1} \sqrt{1-P_{2}}e^{-i\varphi_{2}}e^{i(\xi_{1,0}-\xi_{2,1}-\xi_{f,2})}\nonumber \\
		&   - \sqrt{1-P_1}\sqrt{P_{2}}e^{i\varphi_1}e^{i(\xi_{1,0}+\xi_{2,1}-\xi_{f,2})},
		 \\  
B^{\mathrm{adiab}}_{2}(t_f)&=		 
		 - \sqrt{P_1}\sqrt{P_{2}}e^{i(\xi_{1,0}-\xi_{2,1}+\xi_{f,2})} \nonumber \\
		 & + \sqrt{1-P_1} \sqrt{1-P_{2}}e^{i\varphi_1+\varphi_{2}}e^{i(\xi_{1,0}+\xi_{2,1}+\xi_{f,2})}.
\end{align}

At the end of the interaction with the laser pulse, the probability of transition to the upper level is obtained by projecting Eq. (\ref{Prob_final_factor}) onto the adiabatic wavefunction of the upper level (the conduction band),
\begin{equation}\label{Prob_high}
	C^{\mathrm{adiab}}=
	\begin{pmatrix}
		1 \\ 0
	\end{pmatrix},
\end{equation}
i.e the probability of transition is
\begin{align}\label{eq_final probability}
	\mathcal{P}
	                  &=|C^{\mathrm{adiab}} B^{\mathrm{adiab}}(t_f)|^2 \nonumber \\
			 &= 
			 P_1(1-P_{2}) + P_{2}(1-P_1) \nonumber \\
			 &+ 2\sqrt{P_1 P_{2}(1-P_1)(1-P_{2})}\cos(\varphi_1+\varphi_{2}+2\xi_{2,1}).
	\end{align}
The first two terms in Eq. (\ref{eq_final probability}) are positive definite. The  last term is a
cross-term, which can be both positive or negative. Hence it modulates the transition probability. This term can be associated to interferences. The condition for destructive or constructive interference is when $\cos(\varphi_1+\varphi_2+2\xi_{2,1})$ reaches its minimal negative or maximal positive value, respectively.

\subsection{Interpretation of the numerical EMD}
 Now from Eq. (\ref{eq_pairDensity}), one can express the EMD in the adiabatic-impulse model as the probability of transition defined in Eq. (\ref{eq_final probability}), i.e.
\begin{equation}
f_{s,\mathbf{K}_{\pm}}(\textbf{p}) = \mathcal{P}.
\end{equation}
This relation can serve to guide the interpretation of different features in the EMD, including the presence of interferences and the asymmetry, that are observed in the numerical results depicted in Fig. \ref{fig_momentum_1cycle} (left panels). 
In fact, in the reciprocal space, at the avoided crossing of the adiabatic energies near the Dirac point, electrons under a one-optical cycle laser pulse are promoted from the valence band to the conduction band by laser-induced non-adiabatic transitions following two distinctive pathways at times $t_1$ and $t_{2}$. These two pathways differ in amplitude and phase but lead to electrons with the same final momentum $\mathbf{p}$. This gives rise to interference fringes in momentum space.
 Since the properties of different electron excitation pathways are modified by the CEP-controlled time-dependent field under the envelope, the interference structures in the EMD are directly affected by the CEP. In other words, a modulation  of the temporal waveform by the CEP in time affects the interband coupling in the reciprocal space, which is seen in the modification of the complex interference pattern. 
   This  peak and valley structure can be interpreted as time domain quantum interference, well-known as Landau-Zener-St\"uckelberg interferences (LZSI) \cite{Shevchenko20101}. They have been discussed in the context of graphene in Refs. \cite{PhysRevB.94.125423,Higuchi:2017fk,PhysRevB.93.155434,PhysRevB.88.241112}.

It is interesting to search for conditions when the interband gap is minimal, i.e. when non-adiabatic interband transitions are the most probable according to the adiabatic impulse model \cite{PhysRevB.94.125423}, as described by Eq. (\ref{condition_zero}).
This condition is schematically illustrated in the right panels of Fig. \ref{fig_momentum_1cycle}. For the purpose of illustration, we selected longitudinal momenta where the EMDs have a strong signal (at least in the negative portion of the EMD map), i.e.  $p_x=\pm22.0$ eV for $\phi=0$, $p_x=\pm12.5$ eV for $\phi=\pi/4$ and $p_x=\pm17.0$ eV for $\phi=\pi/2$. The  time evolution of  $p_x + e A(t)$ is represented  in red dashed line for $p_x<0$ and in blue dashed line for $p_x>0$.  When these lines cross the abscissa, the condition of Eq. (\ref{condition_zero}) is satisfied and the electron encounters a minimal interband gap. Under a one cycle laser pulse, there can be up to two solutions to Eq. (\ref{condition_zero}), marked as circles in the plot. These solutions correspond to times $t=t_1,t_2$. From these solutions one can evaluate the instantaneous value of the electric field $E(t_1)$ and $E(t_2)$. At a minimal interband gap, the stronger the instantaneous field amplitude is, the higher the interband coupling  and  the probability of non-adiabatic transition are. 

When the laser pulse is symmetric (panel (b), $\phi=0$), the instantaneous field amplitudes $E(t_{1,2})$ are the same for the longitudinal momenta $p_x<0$ and $p_x>0$. This infers a symmetric EMD as observed in panel (a). In fact,  the non-adiabatic probabilities  obey $P_{1}(-p_{x}) = P_{2}(p_{x})$,  conducting to a symmetric transition probability $\mathcal{P}$, even though there are two distinct quantum pathways.
However, when the laser pulse is asymmetric, the instantaneous field amplitudes $E(t_{1,2})$ for $p_x<0$ are not equivalent to the ones for $p_x>0$. As a consequence, $P_{1}(-p_{x}) \neq P_{2}(p_{x})$ and this causes an asymmetry in the EMD. For $\phi=\pi/4$ (panel (d)) one can see that the instantaneous field amplitude at $t_{1}$ for $p_x=12.5$ eV  is much lower than for $p_x=-12.5$ eV, such that the EMD (panel (c)) has signal mostly on $p_x<0$. For $\phi=\pi/2$, $p_x=17.0$ eV does not even meet the condition for minimal interband gap while it does for its negative counterpart with very high corresponding  instantaneous field amplitude. The EMD is even more asymmetric for $\phi=\pi/2$. 
To sum up, as the CEP controls the asymmetry of the laser pulse, it  imparts an asymmetry to the probability of non-adiabatic transition as a fonction of the longitudinal momentum $p_x$, which is observed in the EMD map.
This asymmetry can in turn be explained in terms of different values of the non-adiabatic Landau-Zener transition probability for different values of the CEP.

\begin{figure}
\includegraphics[height=6.5cm]{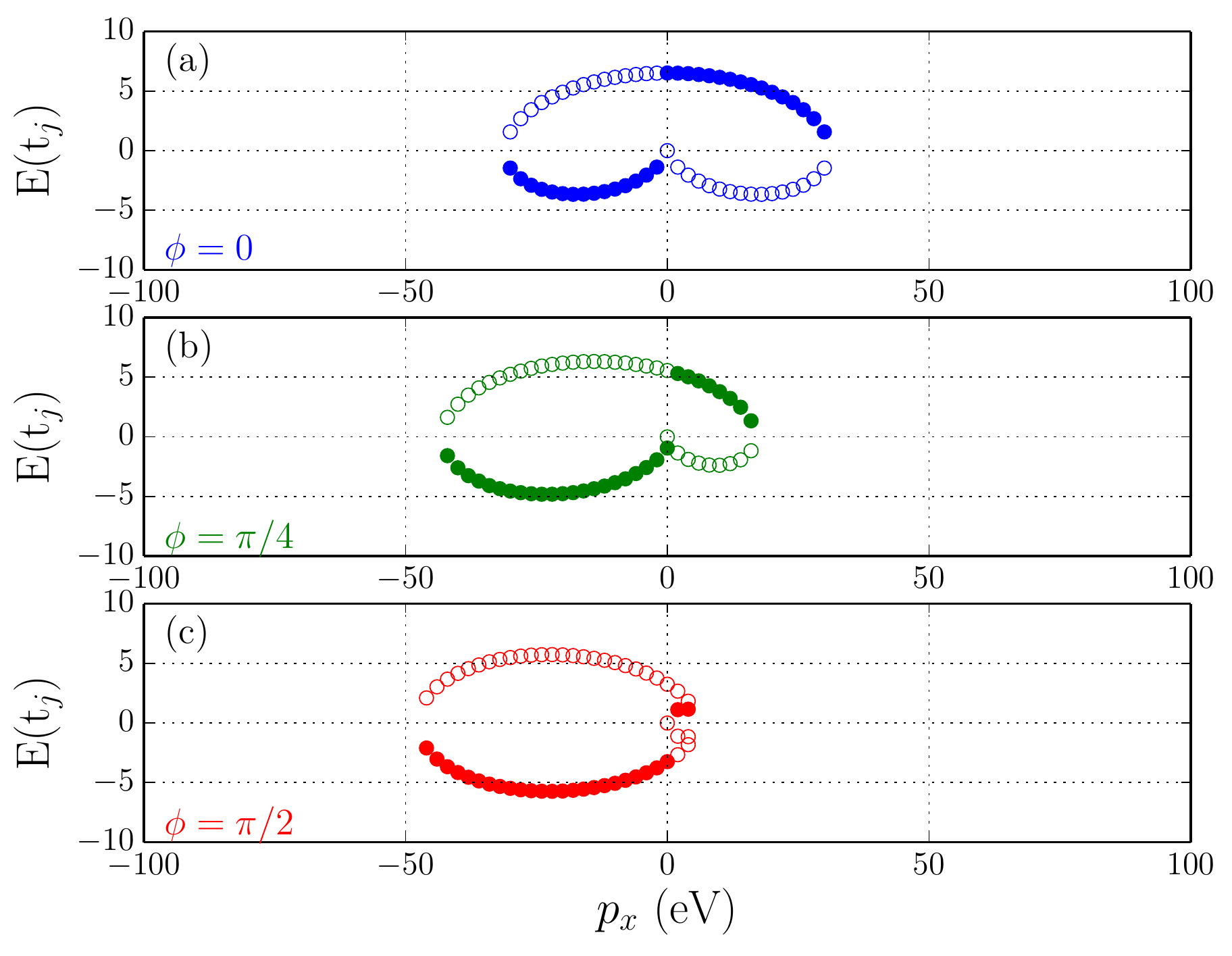}
\caption{Instantaneous field amplitude $E(t_j)$ for $j=1$ (dots) and $j=2$ (circles) when the condition for non-adiabatic transition at minimal interband gap is fulfilled (Eq. (\ref{condition_zero})) for a one-cycle laser pulse ($N=1$), 10 THz,  $10^7$ V/m  and (a) $\phi=0$, (b) $\phi=\pi/4$ and (c) $\phi=\pi/2$.}
\label{fig_p_asym_CEP}
\end{figure}

Following this analysis, one can then extract the instantaneous field amplitudes at  times $t_{1,2}$ satisfying Eq. (\ref{condition_zero}) as a function of the longitudinal momentum $p_x$.
 This is illustrated in Fig. \ref{fig_p_asym_CEP} for (a) $\phi=0$, (b) $\phi=\pi/4$ and (c) $\phi=\pi/2$. The two sets of points (dots and circles) correspond to the two solutions ($j=1$ and $2$, respectively) of Eq. (\ref{condition_zero}). While the case $\phi=0$ gives rise to a symmetric distribution and confirms that $P_{1}(-p_{x}) = P_{2}(p_{x})$, a clear asymmetry is found for the other considered CEPs.   
This asymmetry in the solutions of the non-adiabatic transition condition matches very well the asymmetry observed in the numerical EMD.
One can understand that changing the field strength also affects the range of probability that fulfills the condition for non-adiabatic transitions. 
For stronger fields, the transition probability will be non-vanishing for  larger $|p_x|$ and $|p_y|$, while for weaker fields, the probability will be non-vanishing in a more limited momentum region.  However, it is important to note that the laser intensity must remain in the range of validity of the numerical model used, which considers the dynamics to be constrained to a range of momentum ($|\mathbf{p}|\lesssim 100$ eV) where the Dirac approximation remains valid, i.e. no intervalley coupling. Above such limit, interband non-adiabatic transitions may involve multiple Dirac points  and the tight-binding model is more appropriate for the description of the electron dynamics \cite{PhysRevB.91.045439, Higuchi:2017fk}.

\begin{figure}
\includegraphics[height=6.5cm]{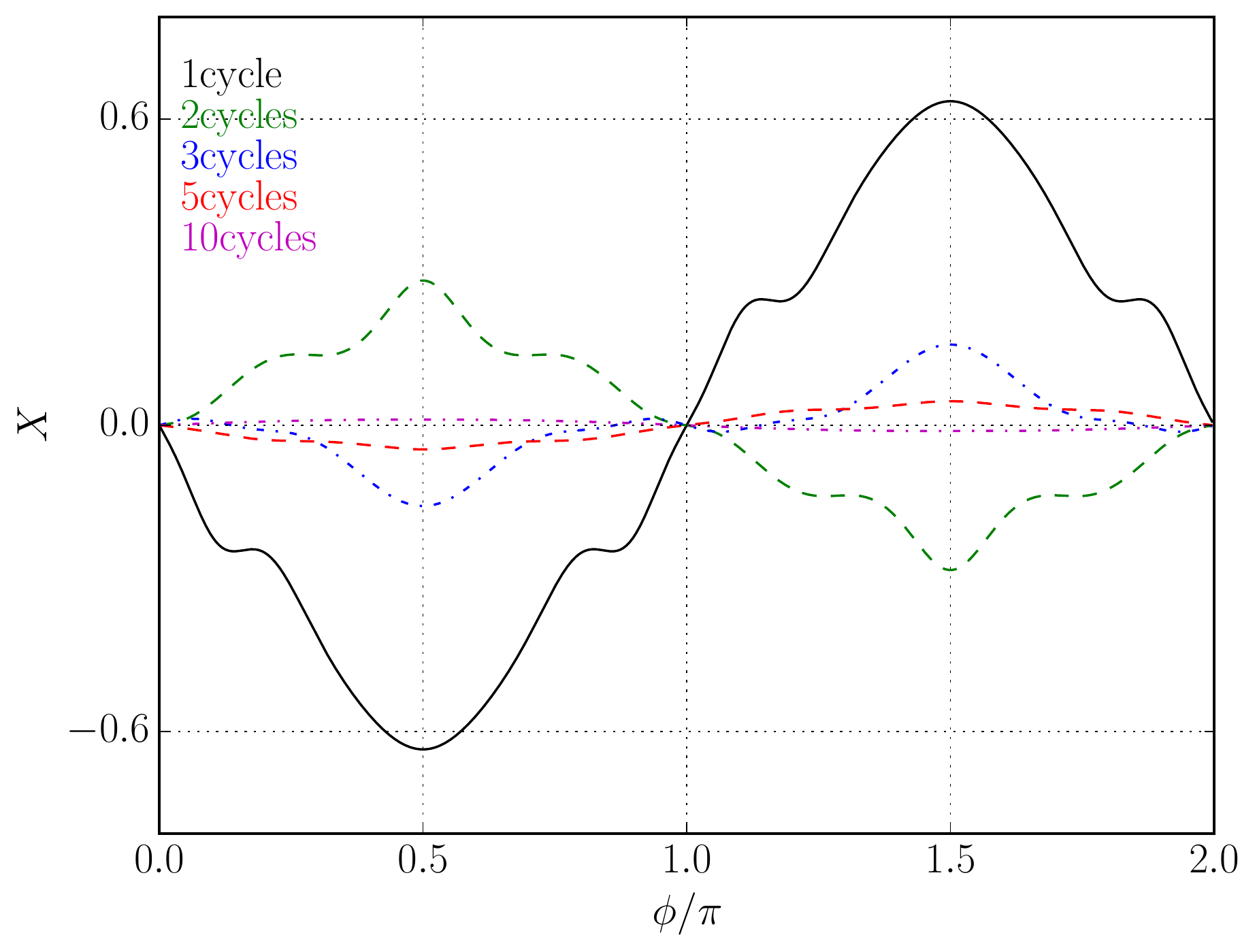}
\caption{Asymmetry parameter $X$ of the integrated EMD as a function of the CEP for laser pulses with 1 (black solid line), 2 (green dashed line), 3 (blue dotted dashed line), 5 (red dashed line), and 10 (magenta dotted dashed line) optical cycles, 10 THz,  $10^7$ V/m.}
\label{fig_X}
\end{figure}

Another way to look at the CEP dependence is to analyze the asymmetry parameter $X$ of the integrated EMD  as a function of the CEP. This parameter is defined as
\begin{equation}\label{eq_asymParam}
X= \frac{P_+-P_-}{P_++P_-},
\end{equation}
where $P_{+}=\int_{0}^{\infty}dp_x\int_{-\infty}^{\infty}dp_y f_{s,\mathbf{K}_{\pm}}(\textbf{p})$ and $P_{-}=\int_{-\infty}^{0}dp_x\int_{-\infty}^{\infty}dp_y f_{s,\mathbf{K}_{\pm}}(\textbf{p})$ are the EMD integrated over positive and negative longitudinal momentum $p_x$, respectively. 
This is a particularly interesting quantity because it determines the direction of the field-induced current in the graphene sample \cite{Higuchi:2017fk,PhysRevLett.116.197401}.
$X$ is displayed in Fig. \ref{fig_X} in black solid line.
One can see that the asymmetry is negative for  $\phi<\pi$, meaning that the momentum map is shifted towards negative values of $p_x$, while the asymmetry is positive for $\phi>\pi$. The asymmetry is the highest around $\phi=\pi/2$ and $3\pi/2$, corresponding to configurations where the vector potential of the laser pulse is the most asymmetric with respect to the centre of the laser pulse envelope, i.e. the maximum of the laser amplitude $E_0$. 
We note that for other laser excitation conditions (e.g. number of half-cycles), the direction of the electron excitation in $p_x$ may change. However a change of $\pi$ in CEP will always reverse the momentum asymmetry with respect to $p_x=0$. The observed momentum asymmetric distribution leads to a residual current after the pulse excitation. For instance, an asymmetry towards positive momenta $k_x>0$ corresponds to a flow of electrons in the positive $x$-direction and a negative residual current, in the $x$-direction \cite{Higuchi:2017fk,PhysRevLett.116.197401}. Anisotropy of conductivity has also been reported in other conditions where graphene is subjected to a dc high-frequency electromagnetic field of linear polarization \cite{Kristinsson:2016qy}.
 
\subsection{Multi-cycle laser pulse}

Although it is currently experimentally possible to generate one-cycle THz laser pulses \cite{Blanchard:07}, we verify if the asymmetry in the EMD caused by variations of the CEP persists for longer laser pulses. 
With a one-cycle pulse, we have shown quantitatively, based on the adiabatic-impulse model, how the EMD is governed by the interband non-adiabatic transitions that occur at each of the two half-cycles. One can understand that with a longer laser pulse, the number of non-adiabatic transitions increases. In fact, we estimated that the number of passages at the adiabatic avoided crossing (quantum pathways) where the non-adiabatic transition condition is fulfilled is $ \sim 2^{2N-1}$, $2N$ being the number of half-cycles. Consequently, the more optical cycles in the laser pulse, the more passages near the minimal interband gap such that  
the interference structure becomes narrower. In addition, while the theoretical interpretation via the adiabatic impulse model becomes prohibitive, it remains straightforward to numerically compute the resulting EMD.  

\begin{figure}
\centering
             %
\includegraphics[height=10cm]{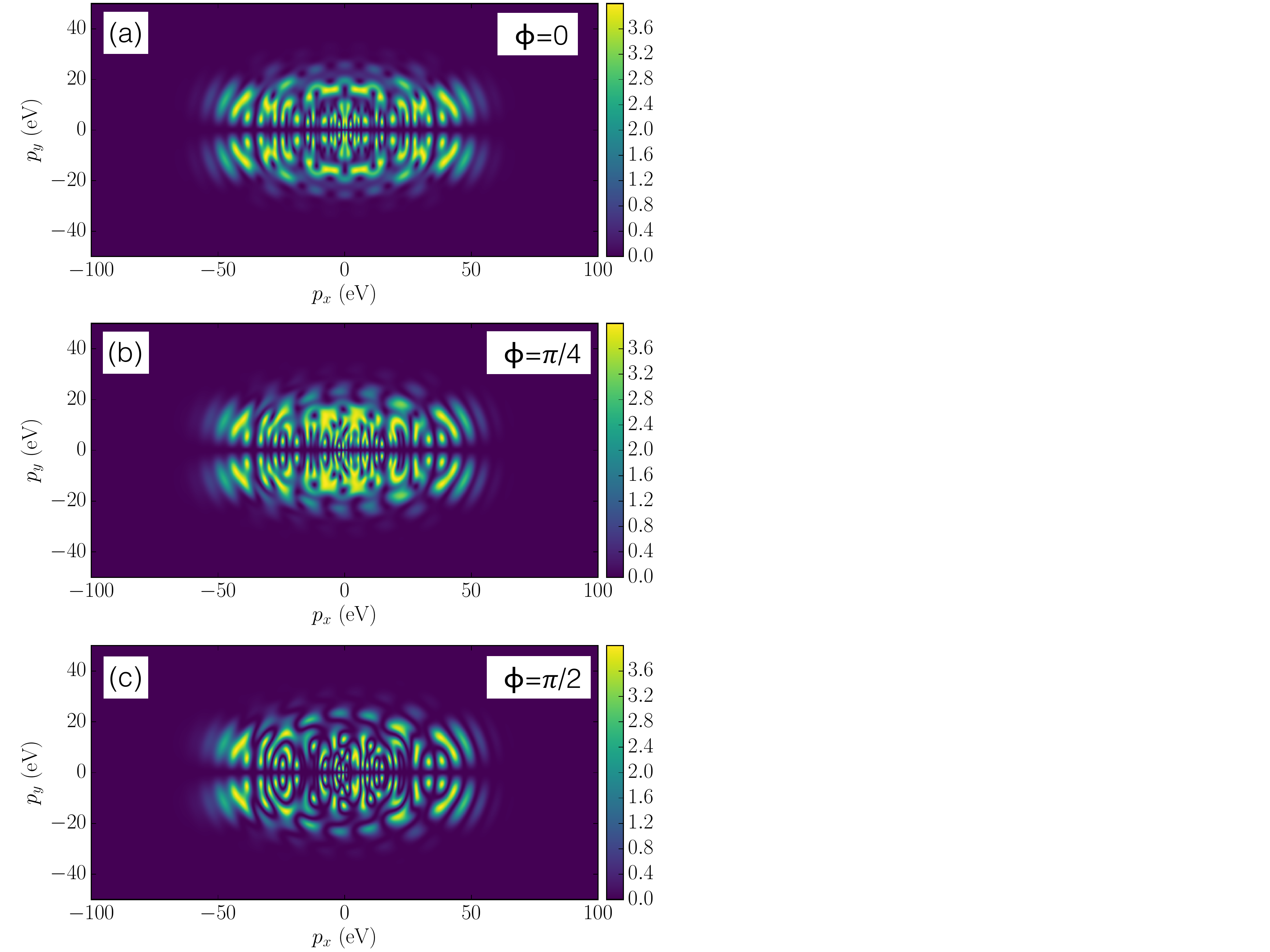}
             \caption{Electron momentum distribution maps in graphene for three different values of CEP (a) $\phi=0$, (b) $\phi=\pi/4$ and (c) $\phi=\pi/2$ for a 5 cycle laser pulse ($N=5$); 10 THz;  $10^7$ V/m. }\label{fig_momentum_5cycles}
\end{figure}

\begin{figure}
\includegraphics[height=6.5cm]{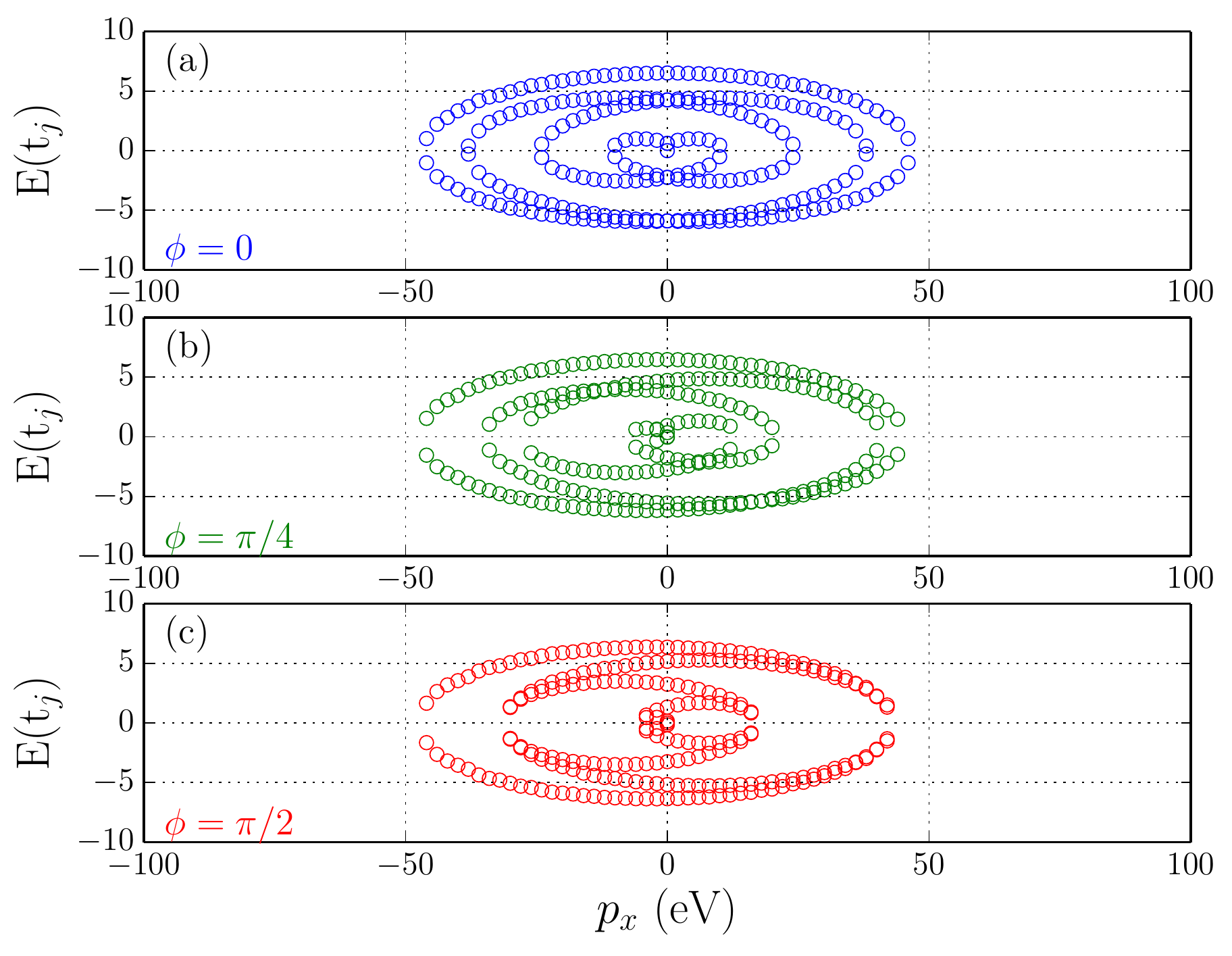}
\caption{Instantaneous field amplitude $E(t_j)$ for $j=1$ to 512 when the condition for non-adiabatic transition at minimal interband gap is fulfilled (Eq. (\ref{condition_zero})) for a 5 cycle laser pulse ($N=5$), 10 THz,  $10^7$ V/m  and (a) $\phi=0$, (b) $\phi=\pi/4$ and (c) $\phi=\pi/2$.}
\label{fig_p_asym_CEP_5cnop}
\end{figure}

The numerical EMD is for instance illustrated in Fig. \ref{fig_momentum_5cycles} with a 5 optical cycles pulse ($N=5$), with the same frequency and field strength as previously used.  
 We notice that the interference pattern in the EMD is indeed more complex as compared to the case of one optical cycle (Fig. \ref{fig_momentum_1cycle}).  According to the adiabatic-impulse model, a total of 512 quantum pathways satisfy the non-adiabatic transition condition. As shown in Fig. \ref{fig_p_asym_CEP_5cnop}, the model predicts solutions of the non-adiabatic transition conditions with a wide lateral momentum distribution from $p_x \in [-50\mathrm{eV},50\mathrm{eV}]$. It also predicts an asymmetry in the distribution at lower lateral momenta for $\phi\ne0$. 
In fact, there is a certain degree of asymmetry in the numerical EMD (Fig. \ref{fig_momentum_5cycles}) for $\phi=\pi/4$ (panel (b)) and $\pi/2$ (panel (c)), while the EMD is perfectly symmetric for $\phi=0$ (panel (a)). 
However, the adiabatic-impulse model  cannot explain all the complex features.
Such features include ring structures. These arise from multiphoton processes: When the energy-level splitting between the valence band and the conduction band matches the energy of an integer number of photons, $k$, of the exciting laser field, then interband transitions can take place by absorption of these $k$ photons \cite{PhysRevB.94.125423,0953-8984-29-3-035501,Gagnon2017}.  

  The asymmetry parameter $X$ is also calculated in Fig. \ref{fig_X} for laser pulses of duration of 2, 3, 5 and 10 optical cycles. As discussed previously, there is still an  inversion with respect to $p_x=0$. In addition, the sign of the asymmetry depends on the number of optical cycles under the pulse envelope. In actuality, under a few-cycle laser pulse of $\phi\ne0$, the highest field amplitude, at the $t=T/2$, is always located on one half-cycle, which is the one dominating the process of  interband non-adiabatic transition. As the number of optical cycles changes from odd to even, the sign of the electric field dominating this half-cycle is reversed. This explains the reversed asymmetry observed in the EMD. The result of such persisting asymmetry, if it could be observed, is a change of direction in the residual current generated as the CEP varies for a given number of optical cycles in a laser pulse.

  However, with  a multi-cycle laser pulse, the effects of the CEP is less compelling for two reasons. First, the asymmetry decreases rapidly with the number of optical cycles and becomes negligible for more than 10 optical cycles. Second, for longer pulses with duration of 100 fs or more, the effect of electron-electron scattering increases. Indeed, as mentioned earlier, the typical scattering time scale is a few tens of femtoseconds \cite{ANDP:ANDP201700038}. The main effect of scattering is to steer the system towards equilibrium with a Fermi-Dirac-like isotropic EMD. In other words, if the pulse is too long, scattering will significantly reduce any anisotropy or interference in the EMD and will thus erase any CEP effect.

 In short, a CEP-tunable one-cycle THz pulse is preferable to a few-cycle pulse for the observation of relativistic-like electron dynamics in monolayer graphene and other Dirac materials, especially since it results in larger and more directional induced currents.



\section{Conclusion}\label{section_conclusion}

To conclude, in this paper we observe a clear signature of the CEP of a one-cycle laser pulse on the laser-induced electron-hole pair production in graphene. 
Using the adiabatic impulse-model, we define conditions for the occurrence of non-adiabatic interband transitions and describe quantitatively how the CEP has an effect both on the number of occurrences of these transitions and on the associated Landau-Zener transition probability.
Successive non-adiabatic transitions give rise to quantum interferences that are seen in the electron momentum distribution. A one-cycle pulse in the THz spectral range is experimentally feasible \cite{Blanchard:07}. This means that our studies benchmark CEP effects that should be experimentally observable with CEP tunable THz radiation.
For a longer pulse, the electron dynamics encounters more non-adiabatic transitions such that the interference pattern in momentum space becomes both more complex and less asymmetric. Our study suggests a way to control the electron dynamics with the CEP in other Dirac materials, as this is not only specific to monolayer graphene.
The observed control can then be looked in a reversed way, where the specific EMD can help to diagnose the laser pulse characteristics. This evokes an alternative way to measure the CEP of ultrashort pulses in the THz regime in interaction with condensed matter, if single-shot measurements of the EMD are possible. Conversely, it also implies that the measurement of an observable requiring an accumulation of many laser shots for statistical purposes will also require the laser pulses to be CEP-stabilized.

\section{Funding Information}

The authors acknowledge Joey Dumont for revision of the manuscript and Philippe Blain for code development.
DG is supported by a FRQNT postdoctoral research scholarship.
Computations were made on the supercomputer Mammouth-Parallel II from Universit\'e de Sherbrooke, managed by Calcul Qu\'ebec and Compute Canada. The operation of this supercomputer is funded by the Canada Foundation for Innovation (CFI), the minist\`ere de l'\'Economie, de la science et de l'innovation du Qu\'ebec (MESI) and  the FRQ-NT.

\appendix

\section{Appendix: expression of free states}
\label{app:free_states}
The explicit expression of the positive and negative energy states is given by
\begin{align}
\label{eq:free_spin1}
u_{s,\mathbf{K}_{+}}(\mathbf{p}) &= 
\cfrac{1}{\sqrt{E_{\mathbf{p}}}}
\begin{bmatrix}
E_{\mathbf{p}} \\
v_{F}(p_{x} + ip_{y} )
\end{bmatrix}, \\ 
\label{eq:free_spin2}
v_{s,\mathbf{K}_{+}}(-\mathbf{p}) &= 
\cfrac{1}{\sqrt{E_{\mathbf{p}}}}
\begin{bmatrix}
v_{F}(p_{x} - ip_{y} )\\
E_{\mathbf{p}} 
\end{bmatrix} ,\\
\label{eq:free_spin3}
u_{s,\mathbf{K}_{-}}(\mathbf{p}) &= \cfrac{1}{\sqrt{E_{\mathbf{p}}}}
\begin{bmatrix}
E_{\mathbf{p}} \\
v_{F}(-p_{x} - ip_{y}) 
\end{bmatrix} ,\\
\label{eq:free_spin4}
v_{s,\mathbf{K}_{-}}(-\mathbf{p}) &= \cfrac{1}{\sqrt{E_{\mathbf{p}}}}
\begin{bmatrix}
v_{F}(-p_{x} + ip_{y} )\\
E_{\mathbf{p}} 
\end{bmatrix},
\end{align}
where $E_{\textbf{p}} = v_{F}|\mathbf{p}|$ is the energy.



 

\end{document}